\documentclass[aps,showpacs,twocolumn,twoside]{revtex4}
\usepackage{epsfig,amssymb,amsmath,amsfonts}
\begin{document}

\def\ket#1{|#1\rangle}
\def\bra#1{\langle#1|}
\def\scal#1#2{\langle#1|#2\rangle}
\def\matr#1#2#3{\langle#1|#2|#3\rangle}
\def\bino#1#2{\left(\begin{array}{c}#1\\#2\end{array}\right)}
\def\ave#1{\langle #1\rangle}
\def\dis#1{\langle\langle #1\rangle\rangle}
\def\uvo#1{\lq #1\rq\ }
\def\uuvo#1{\lq\lq #1\rq\rq}
\def\ave#1{\langle#1\rangle}
\newcommand{\field}[1]{\mathbb{#1}}

\title{Quantum quench influenced by an excited-state phase transition}

\author{
  P. P\'erez-Fern\'andez,$^{1}$
  P. Cejnar,$^{2}$
  J. M. Arias,$^{1}$ 
  J. Dukelsky,$^{3}$
  J. E. Garc\'{\i}a-Ramos,$^{4}$ and 
  A. Rela\~{n}o$^{5}$
 }

 \affiliation{
$^{1}$ Departamento de F\'{\i}sica At\'omica, Molecular y
  Nuclear, Facultad de F\'{\i}sica, Universidad de Sevilla,
  Apartado~1065, 41080 Sevilla, Spain \\
$^{2}$ Institute of Particle and Nuclear Physics, Faculty of
  Mathematics and Physics, Charles University, V Hole{\v s}ovi{\v
  c}k{\'a}ch 2, Prague, 18000, Czech Republic \\
$^{3}$ Instituto de Estructura de la Materia, CSIC, Serrano 123,
E-28006 Madrid, Spain \\
$^{4}$ Departamento de F\'{\i}sica Aplicada, Universidad de Huelva,
  21071 Huelva, Spain \\
$^{5}$ Grupo de F\'{\i}sica Nuclear, Departamento de F\'{\i}sica At\'omica, Molecular y
  Nuclear, Universidad Complutense de Madrid, Av. Complutense s/n, 28040
  Madrid, Spain 
 }

\date{\today}
\begin{abstract}
We analyze excited-state quantum phase transitions (ESQPTs) in three schematic (integrable and nonintegrable) models describing a single-mode bosonic field coupled to a collection of atoms. 
It is shown that the presence of the ESQPT in these models affects the quantum relaxation processes following an abrupt quench in the control parameter. 
Clear cut evidence of the ESQPT effects is presented in integrable models, while in the nonintegrable model the evidence is blurred due to chaotic behavior of the system in the region around the critical energy.
\end{abstract}
\maketitle

\section{Introduction}
\label{int}

Diverse quantum effects in systems depending on external control parameters represent an interesting field of theoretical and experimental investigation.
A lot of recent attention in this field has been focused on two different types of dynamical phenomena, namely on quantum phase transitions and so-called quantum quenches.
These phenomena and their mutual relation are addressed in the present work.

A quantum phase transition (QPT) is a sudden change of the ground-state structure at a certain critical value of the control parameter $\lambda$.
It can be observed as a nonanalytic evolution of the system's energy and wave function induced by an {\em adiabatic} variation of the control parameter $\lambda$ across the quantum critical point $\lambda_{c0}$ at zero temperature.
First discussed in the 1970s \cite{Her76,Gil78,Gil81}, the QPT phenomena become very important in the context of solid state physics \cite{Son97,Sac99,Voj03} as well as in nuclear and many-body physics---see, e.g., recent reviews \cite{Cast09,Cej10}.

A quantum quench (QQ) represents an abrupt, {\em diabatic} change $\lambda_1\to\lambda_2$ of the control parameter followed by a system-specific quantum relaxation process.
Pioneering theoretical works in this field appeared already in the late 1960s \cite{Bar69}, but a really rapid growth of interest was triggered by experimental studies at the beginning of this millennium \cite{Gre02}.
For an extensive list of QQ-related references see Ref.~\cite{Par09}.

Not surprisingly, the QPT and QQ effects can be mutually related.
If the initial state before a quench coincides with the ground state near a QPT, the dynamics after the quench depends substantially on whether the parameter change does or does not bring the system to the other quantum phase, or eventually to the narrow quantum critical region between the phases \cite{Par09,Sen04,Sil08,Ven10}.

In this paper, we discuss the quench-induced dynamics in connection with a novel concept related to quantum criticality---a so called excited-state quantum phase transition (ESQPT) \cite{Cej06,Cap08,Cej08}.
This phenomenon represents a nonanalytic evolution of individual {\em excited} states in the system with a variable control parameter.
So far, such effects have been studied mostly in integrable systems with one effective degree of freedom (one dimensional configuration spaces), showing a singularity of classical dynamics at a certain energy \cite{Ley05,Rei05,Hei06,Rib07,Rel08,Per08,Kan09,Fig10}, but they seem to exist in a much richer variety of incarnations.

The ESQPTs can be viewed as a reinterpretation of thermal phase transitions in the microcanonical language.
The key step is a scaling of energy and other observables by a suitably defined size parameter $\aleph$.
Let $\ave{\bullet}_T$ stands for a thermal average of the quantity in brackets at temperature $T$.
If the thermal fluctuation $\ave{\Delta{\cal E}^2}_T\equiv\ave{{\cal E}^2}_T-\ave{{\cal E}}^2_T$ of the scaled energy ${\cal E}\equiv E/\aleph$ vanishes in the thermodynamic limit $\aleph\to\infty$ (often synonymous with the classical limit $\hbar\to 0$), a thermal phase transition at temperature $T=T_c$ becomes localized at a sharp value ${\cal E}_c=\ave{{\cal E}}_{T_c}$ of the scaled energy.
This shows up as an anomalous (nonanalytic) \uuvo{flow} of energy levels ${\cal E}_i(\lambda)$ through the boundary ${\cal E}_c(\lambda)$ defining the locus of ESQPT points \cite{Cej06,Cap08,Cej08,Ley05,Hei06}.

Besides the shape of individual ${\cal E}_i(\lambda)$ curves, the excited-state transition affects also the dependence of the level density $\rho({\cal E})$ on the scaled energy.
The function $\rho({\cal E})$ at ${\cal E}={\cal E}_c$ shows a singularity whose type depends on the underlying thermal phase transition and/or on the corresponding anomaly in the associated classical phase space \cite{Cap08,Ley05,Rei05}.
Since the singularity influences quantum relaxation processes in the critical region, we can anticipate a major impact of the ESQPT on the QQ-induced dynamics in those cases in which the energy distribution after the quench is centered near ${\cal E}_c$.

To study the above-formulated conjecture, we use three simple models.
They describe a single-mode bosonic field interacting with an algebraic subsystem, which is based either on the SU(1,1) or on the SU(2) dynamical algebra.
The SU(1,1) model may serve as a toy for the description of formation and dissociation of diatomic molecules and bosonic atoms \cite{Tik08}. 
The SU(2) Hamiltonian represents either the well-known Dicke \cite{Dic54} or Jaynes-Cummings \cite{Jay63} (Tavis-Cummings \cite{Tav68}) models of quantum optics, or may alternatively describe an interacting mixture of diatomic molecules and fermionic atoms \cite{Tik08}. 
In this work we will adopt the former interpretation.
While the SU(1,1) model and the SU(2)-based Jaynes-Cummings model are integrable, the SU(2)-based Dicke model is not.

All the three models show a rather similar phase structure.
First, considering the ground-state properties, it turns out that an increasing strength of interaction between the bosonic field and the algebraic subsystem drives the entire system to a quantum critical point where the ground state abruptly changes its form.
We then show that this QPT is in all three models followed by a chain of ESQPTs and demonstrate that these have a strong impact on the character of relaxation dynamics after some fine-tuned quantum quenches---namely those leading the system to a narrow region around the critical excitation energy.
The type of the ESQPT and its QQ signatures depends on the dimensionality of the model: they are of the strongest type (see below for a precise definition of this concept) for the SU(1,1) and the SU(2) Jaynes-Cummings Hamiltonians, and of a softer type for the SU(2) Dicke Hamiltonian.
We will also see that breaking of integrability in the latter model blurs the effects of criticality in the quench dynamics.
We anticipate the same general trend also in more complex situations.

The plan of the paper is the following:
In Sec.~\ref{mod} we describe the models, and analyze in Sec.~\ref{PTs} their classical and phase-transitional properties, particularly those related to excited states.
Sec.~\ref{dyna} collects the results on quantum quenches.
We introduce a general concept of a critical quench, driving the system into the ESQPT region, and continue to more specific numerical results for the models employed.
Sec.~\ref{con} brings a brief summary and outlook.

\section{Models}
\label{mod}

\subsection{Algebraic structure}
\label{alg}

\begin{figure}
\includegraphics[width=0.7\linewidth]{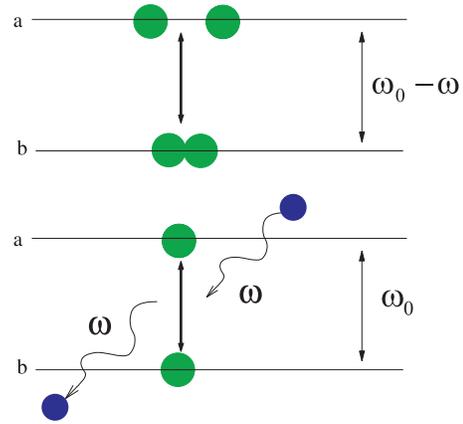}
\caption{A schematic representation of the models used. The SU(1,1) model (the upper panel) may describe the coexistence of two-atom molecules (lower level) with dissociated atoms (upper level). The SU(2) model (the lower panel) describes the interaction of a single-mode radiation field with an array of two-level atoms.}
\label{f_scheme}
\end{figure}

Below we will investigate quantum quenches in two simple models, both describing a system composed of two interacting parts:
(i) a single bosonic mode given by creation and annihilation operators $b^\dag$ and $b$, therefore described by the Heisenberg-Weyl algebra HW(1), and
(ii) a system represented by pseudospin operators  $J_\pm=J_x\pm iJ_y$ and $J_0=J_z$ satisfying commutation relations of the SU(2) algebra,
\begin{equation}
[J_0,J_\pm]=\pm J_\pm,\ [J_+,J_-]=2J_0,
\end{equation}
or by analogous operators $K_\pm=K_x\pm iK_y$ and $K_0=K_z$ satisfying commutation relations of the SU(1,1) algebra
\begin{equation}
[K_0,K_\pm]=\pm K_\pm,\ [K_+,K_-]=-2K_0.
\end{equation}
The SU(2) or SU(1,1) algebras will be realized more specifically in terms of fermionic or bosonic operators.
The complete dynamical algebra is ${\rm HW}(1)\,\otimes\,{\rm SU}(\bullet)$, where the bullet stands for a specification of the respective special unitary algebra, but in the following we will use just abbreviated names SU(2) and SU(1,1) for the two models.
A schematic representation of both models is given in Fig.~\ref{f_scheme}.

The Hilbert space of the coupled system is identified with the tensor product $\field{H}=\field{H}^{\rm (i)}\otimes\field{H}^{\rm (ii)}$, where $\field{H}^{\rm (i)}$ is the space of HW(1) spanned by the set of basis vectors $\ket{N_b}$, with $N_b=0,1,\dots$ denoting the number of $b$-bosons, and $\field{H}^{\rm (ii)}$ coincides with the space associated with one of the irreducible representations (irreps) of the groups SU(2) or SU(1,1).
The irreps are classified by the eigenvalues $c^{(2)}_{\rm SU(\bullet)}$ of the respective second-order Casimir invariant,
\begin{eqnarray}
C^{(2)}_{\rm SU(2)}&=&
J_x^2+J_y^2+J_z^2\,,\\
C^{(2)}_{\rm SU(1,1)}&=&
K_x^2+K_y^2-K_z^2,
\end{eqnarray}
which are parametrized as $c^{(2)}_{\rm SU(2)}=j(j+1)$ (with $j$ integer or half-integer) and $c^{(2)}_{\rm SU(1,1)}=-k(k-1)$ (with $k>0$ known as the Bergmann index).
The irreps are finite-dimensional in the SU(2) case (compact group) and infinite-dimensional in the SU(1,1) case (noncompact group).
The respective basis states $\ket{j,m}$ (with $m=-j,-j+1,\dots,+j$ being the eigenvalue of $J_0$) and $\ket{k,n}$ (with $n=0,1,\dots$ enumerating the eigenvalues $k+n$ of $K_0$) are generated from the lowest state $\ket{j,-j}$ and $\ket{k,0}$ by consecutive actions of the rising operators $J_+$ and $K_+$, respectively.

\subsection{SU(1,1) model}
\label{1d}

Since the group generated by the SU(1,1) algebra is noncompact, its irreps are infinite dimensional, the generators being expressible through creation and annihilation operators $a^\dag,a$ of another type of bosons.
For instance, a single boson pair realization reads as
\begin{equation}
K_+=\tfrac{1}{2}(a^\dag)^2,\ K_-=\tfrac{1}{2}\,a^2,\ K_0=\tfrac{1}{2}\left(a^\dag a+\tfrac{1}{2}\right).
\label{su11g}
\end{equation}
Alternatively, one can use some other boson pair realizations (e.g., with two kinds of bosons), which together with Eq.~(\ref{su11g}) constitute various forms of the Schwinger representation of the SU(1,1) algebra.
Note that in Sec.~\ref{clas} we will also introduce the Holstein-Primakoff bosonic representation.

To construct the Hamiltonian, we assume the simplest realization (\ref{su11g}).
In this case, there are just two irreps, one with $k=\frac{1}{4}$ and the other with $k=\frac{3}{4}$.
Their respective Hilbert spaces are spanned by vectors $\ket{N_a}$ containing even and odd numbers of $a$-bosons.
The interaction between $a$- and $b$-bosons is considered such that the creation of one $b$-boson leads to the destruction of a pair of $a$-bosons and vice versa.
Such a model can schematically describe, for example, the formation and dissociation of two-atom molecules \cite{Tik08}.
The total Hamiltonian reads as
\begin{equation}
H^{(1)}=
\omega_0 K_0 +
\omega b^\dag b +
\frac{\lambda}{\sqrt{M^{(1)}}}\biggl[bK_++b^\dag K_-\biggr]\,,
\label{hsu11}
\end{equation}
where $\lambda/\sqrt{M^{(1)}}\geq 0$ is a scaled coupling parameter (the meaning of $M^{(1)}$ will be explained below) and $\omega,\omega_0$ stand for single-particle energies (we set $\hbar=1$).

For each of the SU(1,1) irreps (classified by the quantum number $k$), there are two commuting operators (quantum degrees of freedom) which determine the basis in the whole Hilbert space of physical states: one is associated with the number of $b$-bosons, $N_b=b^\dag b$, the other with $K_0$, or equivalently with the number of $a$-bosons $N_a=a^\dag a$.
At the same time, there exist two different integrals of motions: one is the energy $H$ and the other one can be written in the form
\begin{equation}
M^{(1)}=2N_b+N_a-\tfrac{4k-1}{2}=2(N_b+K_0-k)
\,.
\label{em}
\end{equation}
The value of $M^{(1)}\geq 0$ is always even, $M^{(1)}/2$ counting the number of $b$-bosons plus the number of $a$-boson pairs.
The conservation of $M^{(1)}$ implies that the Hamiltonian (\ref{hsu11}) represents an integrable system, which for each fixed value of $M^{(1)}$ can be associated with an effective one-dimensional configuration space (one quantum degree of freedom).

In the following we will assume $\omega_0>\omega$ for the SU(1,1) model.
This means that the $\lambda=0$ ground state can be identified with a molecular condensate with no pair of atoms.
It has the form $\ket{N_b=M^{(1)}/2}\otimes\ket{k,0}$, where the first term represents a state with a maximal number of $b$ bosons and the second one stands for the lowest weight SU(1,1) state with the minimal value of $N_a=0$ ($N_a=1$) for $k=\frac{1}{4}$ ($k=\frac{3}{4}$).
However, for sufficiently large values of the coupling parameter $\lambda$ the interaction between the molecules and atomic pairs supports a more balanced distribution of the expectation values $\ave{N_a}$ and $\ave{2N_b}$.
With an increasing size of the system the crossover between the two types of the ground-state structure is getting sharper and in the infinite-size limit, $M^{(1)}\to\infty$, it becomes a phase transition.
The calculation of the critical value of the interaction strength $\lambda^{(1)}_{c0}$ will be presented in Sec.~\ref{qpt}.

\subsection{SU(2) model}
\label{2d}

The SU(2) algebra yields a compact group with finite-dimensional irreps.
Its generators can therefore be constructed from fermionic operators.
For instance, they can be associated with an array of spin-$\frac{1}{2}$ particles (or two-level atoms) located on $2j$ sites:
\begin{eqnarray}
J_+=\sum_{i=1}^{2j} a_{\uparrow i}^\dag a_{\downarrow i},\
J_-=\sum_{i=1}^{2j} a_{\downarrow i}^\dag a_{\uparrow i},\nonumber\\
J_0=\tfrac{1}{2}\sum_{i=1}^{2j}\left(a_{\uparrow i}^\dag a_{\uparrow i}-a_{\downarrow i}^\dag a_{\downarrow i}\right).
\label{su2g}
\end{eqnarray}
Here, $a_{\uparrow i}^\dag$ or $a_{\uparrow i}$ and $a_{\downarrow i}^\dag$ or $a_{\downarrow i}$ create or annihilate spin-up and spin-down states of the fermion on site $i$ and the ladder operators $J_\pm$ describe spin flips along the array.
Alternatively, one can use fermion pair realizations of the SU(2) algebra (with $J_\pm$ creating and annihilating a pair of fermions), or the Schwinger or Holstein-Primakoff types of bosonic realizations with truncated Hilbert spaces (the latter bosonic realization will be discussed in Sec.~\ref{clas}).
Depending on the specific realization, the model can receive different physical interpretations.
Below we will implicitly consider the realization (\ref{su2g}), which may schematically describe interactions of single-frequency photons with two-level atoms in maser-like systems.

The Hamiltonian is taken in either of the following forms,
\begin{eqnarray}
H^{(2)}=\omega_0 J_0 +\omega b^\dag b +
\frac{\lambda}{\sqrt{M^{(2)}}}\biggl[bJ_++b^\dag J_-\biggr],
\qquad\
\label{hsu2}\\
H^{(3)}=\omega_0 J_0 +\omega b^\dag b +
\frac{\lambda}{\sqrt{M^{(3)}}}\biggl[(b+b^\dag)(J_-+J_+)\biggr],
\label{hsu2ni}
\end{eqnarray}
where $\lambda/\sqrt{M^{(2)}}$ or $\lambda/\sqrt{M^{(3)}}$ is a properly scaled coupling parameter ($\lambda\geq 0$) and $\omega,\omega_0$ two single-particle energies.
The Hamiltonian $H^{(2)}$ is known as the Jaynes-Cummings \cite{Jay63} or Tavis-Cummings model \cite{Tav68}, while the Hamiltonian $H^{(3)}$ is referred to as the Dicke model \cite{Dic54}.
Note that a so-called rotating-wave approximation of $H^{(3)}$ leads to the simpler Hamiltonian $H^{(2)}$.

The Jaynes-Cummings Hamiltonian (\ref{hsu2}) is very similar to that of Eq.~(\ref{hsu11}).
It conserves the quantity
\begin{equation}
M^{(2)}=2(N_b+J_0+j)
\,,\label{emm}
\end{equation}
analogous to Eq.~(\ref{em}), and therefore corresponds to an integrable system described effectively by a one-dimensional configuration space.
(The full model has again two degrees of freedom, associated with commuting operators $N_b$ and $J_0$.)
The Dicke model violates the conservation of $M^{(2)}$, but it still conserves the parity $\Pi=(-1)^{M^{(2)}/2}$ labeling individual eigenstates.
In this case, the size parameter is taken as $M^{(3)}=4j$, which is the total number of fermionic states (twice the number of sites).

As in the SU(1,1) case, the ground states of both the SU(2) Hamiltonians change their nature suddenly as the coupling strength $\lambda$ increases above a certain value, the transition having a critical character in the infinite-size limit, $M^{(2)},M^{(3)}\to\infty$.
For the Hamiltonian $H^{(2)}$ from Eq.~(\ref{hsu2}), we will assume $\omega>\omega_0$, identifying the $\lambda=0$ ground state with a photon vacuum, $N_b=0$, combined with a maximally excited state of the atom array: $J_0=\frac{1}{2}M^{(2)}-j$ (below we set $M^{(2)}=4j$ so that $J_0=+j$ at $\lambda=0$).
At the critical coupling strength, this structure eventually changes into a state with $\ave{N_b}>0$, in which a part of energy is transferred from atoms to the photon field.
It should be stressed that here we are running the model in a nonstandard regime, taking into account only a finite set of states with a single fixed value of $M^{(2)}$.
This is in contrast to the rotating-wave approximation of the Dicke model, for which one usually considers the infinite spectrum with all values of $M^{(2)}$.

For the Dicke Hamiltonian $H^{(3)}$ in Eq.~(\ref{hsu2ni}), we set $\omega_0=\omega$, which corresponds to the resonance absorption and emission of photons by the atoms.
The $\lambda=0$ ground state has the form $\ket{N_b=0}\otimes\ket{j,-j}$, describing a photon vacuum and an unexcited array of atoms (recall that for the Dicke model $N_b+J_0$ is not conserved).
For a sufficiently strong interaction between matter and light, the ground state flips to a form with $\ave{J_0}>-j$ and $\ave{N_b}>0$, showing a macroscopic excitation of both subsystems \cite{Hep73,Ema03}.
This may be considered as a toy example of the maser phase transition.

\subsection{Numerical solution}
\label{nume}

The model Hamiltonians from Eqs.~\eqref{hsu11}, \eqref{hsu2} and \eqref{hsu2ni} can be diagonalized numerically in an appropriate basis. 
Due to the tensor product structure of the Hilbert space, the basis is naturally chosen in the form $\ket{N_b}\otimes\ket{k,n}$ for the SU(1,1) model, and $\ket{N_b}\otimes\ket{j, m}$ for both SU(2)-based models, where $\ket{N_b}\in\field{H}^{\rm (i)}$ stands for a state with a given number of $b$ bosons while $\ket{k,n},\ket{j,m}\in\field{H}^{\rm (ii)}$ are basis vectors of the respective SU(1,1) or SU(2) irreps.
Recall that in the SU(1,1) case, the link of $\ket{k,n}$ with the states $\ket{N_a}$, counting the number of $a$ bosons, is achieved via setting $k=\frac{1}{4}$ or $\frac{3}{4}$ for the even- or odd-$N_a$ irreps, respectively, and $n=\frac{1}{2}(N_a-\frac{4k-1}{2})$.
Matrix elements of individual Hamiltonian terms in these bases can be easily calculated from the known action of the $b^\dag$ and $b$ operators on the vectors $\ket{N_b}$ and the action of \{$K_+,K_-,K_0$\} or \{$J_+,J_-,J_0$\} operators on vectors $\ket{j,m}$ or $\ket{k,n}$.

For both SU(1,1) and SU(2) integrable models, the basis includes a finite set of vectors.
These are determined by the chosen values of the size parameters $M^{(1)}$ or $M^{(2)}$, which permit only a finite number of combinations of $(N_b,n)$ or $(N_b,m)$ satisfying Eqs.~(\ref{em}) and (\ref{emm}), respectively.
Thus the corresponding Hamiltonian matrices are finite and the diagonalization is just a routine problem.

The nonintegrable Dicke model, on the other hand, has no conservation-dictated constraint on the allowed combinations of basis vectors. 
Its basis is therefore infinite and must be numerically truncated for $N_b>N_{\rm trunc}$, making the convergence tests of the diagonalization outputs an important issue.
Speaking qualitatively, the truncation with a fixed $N_{\rm trunc}$ can be safely done in the low-energy part of the spectrum only for a sufficiently small interaction strength $\lambda$ between the $b$ bosons (photons) and atoms.
Indeed, we know that for $\lambda=0$, the Dicke Hamiltonian is diagonal in the $\ket{N_b}\otimes\ket{j, m}$ basis, the states with increasing $N_b$ being associated with increasingly high excitations.
Therefore, for moderate values of $\lambda$ the states with high photon numbers are only weakly admixed to the low-lying energy eigenstates of the system.
As $\lambda$ increases, the cutoff parameter $N_{\rm trunc}$ must increase accordingly for a given low-energy portion of the spectrum to be well reproduced.
We stress that the results presented below were tested for stability against the change of $N_{\rm trunc}$ (see Sec.~\ref{renoin}).

The setting of the model parameters, as used in the calculations below, and some important model-specific values are summarized in Tab.~\ref{sumtab} (some of the symbols will be explained later).

\begin{table}[t]
\begin{tabular}{| l || c c |c | c|}
\hline
 & \multicolumn{2}{|c|}{SU(1,1)} & \multicolumn{2}{|c|}{SU(2)} \\
 \cline{2-5}
 & \multicolumn{2}{|c|}{$n=1$} & $n=2$ & $n=3$ \\ 
 & $N_a$ even & $N_a$ odd & Jayn.-Cumm. & Dicke \\
\hline\hline
$\omega,\omega_0$ & \multicolumn{2}{|c|}{$\omega_0-\omega=1=\omega$} & $\omega-\omega_0=1=\omega_0$ & $\omega=1=\omega_0$ \\
\hline 
$M^{(n)}$  & $2N_b$+$N_a$  & $2N_b$+$N_a$$-$1  & $4j$ & $4j$ \\
\hline
$R^{(n)}$ & $\frac{1}{2M^{(1)}}$ & $\frac{3}{2M^{(1)}}$ & $\frac{1}{2}$ & $\frac{1}{2}$ \\ 
\hline
$\lambda_{c0}^{(n)}$ & \multicolumn{2}{|c|}{$\frac{1}{\sqrt{2}}$} & $\frac{1}{\sqrt{2}}$ & $\frac{1}{\sqrt{2}}$ \\
\hline
${\cal E}_c^{(n)}$ & \multicolumn{2}{|c|}{$\frac{1}{2}$} & $\frac{1}{4}$ & $-\frac{1}{4}$ \\ 
\hline
\end{tabular}
\caption{Summary of parameter setting for the three models employed: frequencies $\omega,\omega_0$, the size parameter $M^{(n)}$, a parameter $R^{(n)}$ in the potential energy, the QPT critical parameter $\lambda_{c0}^{(n)}$, and the ESQPT critical scaled energy ${\cal E}_c^{(n)}$.}
\label{sumtab}
\end{table}

\section{Phase transitions}
\label{PTs}

\subsection{Classical limit}
\label{clas}

The above Hamiltonians are specimens of a rather large general class of systems---namely those described by finite algebraic models \cite{Bar88}.
For these models, the relevant observables are constructed in terms of a finite set of generators $G_i$ closing a dynamical algebra $[G_i,G_j]=\sum_kc_{ijk}G_k$ with structure constants $c_{ijk}$.
The corresponding systems have a finite number of degrees of freedom and their thermodynamic (infinite size) limit coincides with the classical limit $\hbar\to 0$.
To see this, recall that the thermodynamic limit is generally achieved for asymptotic values of a properly defined size parameter $\aleph$ such that thermal fluctuations of a scaled Hamiltonian ${\cal H}=H/\aleph$ vanish with $\aleph\to\infty$.
In algebraic systems, this parameter needs to be introduced on the level of individual generators, via scaled generators ${\cal G}_i\equiv G_i/\aleph^\kappa$ (with $\kappa>0$) whose substitution into the Hamiltonian $H(G_i)$ should be consistent with the definition of ${\cal H}$, thus $H(G_i/\aleph^\kappa)=H(G_i)/\aleph$.
In fact, this is why the size parameter $\aleph\equiv M^{(n)}$ was included into the effective coupling constant $\lambda/\sqrt{\aleph}$ of the above Hamiltonians $H^{(n)}$ ($n=1,2,3$).
The known commutation relations for the bare generators $G_i$ then ensure that the scaled generators ${\cal G}_i$ yield vanishing commutators in the $\aleph\to\infty$ limit, $[{\cal G}_i,{\cal G}_j]\to 0$, which constitutes the classical behavior of the correctly scaled observables.

A general method for approaching the classical limit in finite algebraic models is based on coherent states \cite{Per86,Zha90}.
These for the above-described composite systems are naturally considered in the form of a tensor product $\ket{\zeta}\otimes\ket{\xi}$, where $\ket{\zeta}\propto e^{\zeta b^\dag}\ket{0}$ (with $\zeta\in\field{C}$) is the HW(1) coherent state of the subsystem (i), and $\ket{\xi}$ is a yet unspecified SU(1,1) or SU(2) coherent state of the subsystem (ii) \cite{Kur89,Cas09}.
The latter states can be taken in several alternative forms, depending on a concrete realization of the two algebras.
One possibility is to use $\ket{\xi}\propto e^{\xi K_+}\ket{k,0}$ or $\ket{\xi}\propto e^{\xi J_+}\ket{j,-j}$ (with $\xi\in\field{C}$) and associate the classical limit with $j\to\infty$ or $k\to\infty$.
The corresponding phase space of subsystem (ii) is then identified with a 2D surface of constant positive or negative curvature, which is the sphere $j_x^2+j_y^2+j_z^2=1$ in the SU(2) case ($j_i=J_i/j$) or a two-sheet hyperboloid $k_x^2+k_y^2-k_z^2=-1$ in the SU(1,1) case ($k_i=K_i/k$).

We do not directly follow this path, partly because in our case the value of the SU(1,1) invariant is fixed to $k=\frac{1}{4}$ or $\frac{3}{4}$, so we would not be able to keep the same treatment in both SU(1,1) and SU(2) systems.
Instead, we employ the Holstein-Primakoff transformation of both algebras onto a bosonic field $c^\dag,c$, which reads as
\begin{eqnarray}
K_+=c^\dag(2k+c^\dag c)^{\frac{1}{2}},\quad
K_-=(2k+c^\dag c)^{\frac{1}{2}}c,
\nonumber\\
K_0=c^\dag c+k,
\qquad\qquad\qquad
\label{su11map}
\end{eqnarray}
in the SU(1,1) case (with $c^\dag c\equiv N_c\geq 0$) and
\begin{eqnarray}
J_+=c^\dag(2j-c^\dag c)^{\frac{1}{2}},\quad
J_-=(2j-c^\dag c)^{\frac{1}{2}}c,
\nonumber\\
J_0=c^\dag c-j,
\qquad\qquad\qquad
\label{su2map}
\end{eqnarray}
in the SU(2) case (with $0\leq N_c\leq 2j$).
In this way, we obtain a mapping of the original ${\rm HW}(1)\otimes{\rm SU}(\bullet)$ dynamical algebra onto a new algebra associated with both $b$ and $c$ types of bosons, which is then analyzed with the aid of bosonic coherent states $\ket{\zeta,\xi}\propto e^{\zeta b^\dag+\xi c^\dag}\ket{0}$, where $\ket{0}$ is a common vacuum of both $b$ and $c$ bosons.

For Hamiltonians (\ref{hsu11}) and (\ref{hsu2}), the new dynamical algebra can be identified with the algebra ${\rm U(2)}\equiv\left\{ b^\dag b, c^\dag c, b^\dag c, c^\dag b \right\}$, since in both these cases the total number of bosons $N=N_b+N_c$ is conserved.
The size parameters $M^{(1)}$ and $M^{(2)}$ introduced above coincide both with the value $2N$.
On the other hand, for the nonintegrable Hamiltonian (\ref{hsu2ni}) the new dynamical algebra can be identified with HW(2), the Heisenberg-Weyl algebra of $b$ and $c$ bosons.
The total number of bosons is not conserved and therefore the only sensible size parameter is the value of $j$ (for consistency reasons we have chosen $M^{(3)}=4j$), which measures the size of the subsystem (ii).

The method proceeds via evaluating the expectation value 
$\matr{\zeta,\xi}{H^{(n)}}{\zeta,\xi}/M^{(n)}$ for any of the above Hamiltonians $H^{(n)}$ (with $n=1,2,3$) in the coherent states of $b$ and $c$ bosons.
The result can be gained directly by substituting
\begin{equation}
\frac{c}{\sqrt{M^{(n)}}}=\frac{x+ip}{\sqrt{2}}\,,\qquad \frac{b}{\sqrt{M^{(n)}}}=\frac{y+iq}{\sqrt{2}}\,,
\label{ani}
\end{equation}
and the Hermitian conjugate expressions for $c^\dag$ and $b^\dag$ into the scaled Hamiltonian ${\cal H}^{(n)}=H^{(n)}/M^{(n)}$ .
Note that in Eq.~(\ref{ani}) we define coordinates $x,y$ and the associated momenta $p,q$, respectively, which satisfy canonical commutation relations $[x,p]=[y,q]=i/M^{(n)}$ with $1/M^{(n)}$ playing the role of Planck constant.

For $M^{(n)}\to\infty$, the coordinate and momentum operators can be treated as commuting variables.
We obtain a general form
\begin{equation}
{\cal H}^{(n)}={\cal H}^{(n)}_0+\lambda {\cal H}'^{(n)}
\label{hhh}
\end{equation}
for the scaled classical Hamiltonian.
The first term,
\begin{equation}
{\cal H}^{(n)}_0=-\frac{R^{(n)}\omega_0}{2}+\frac{\omega_0}{2}(p^2+x^2)+\frac{\omega}{2}(q^2+y^2)
\,,
\label{h0n}
\end{equation}
which describes the system with $\lambda=0$, is common to all three models $n=1,2,3$, with the additive constant expressed as $R^{(1)}=2k/M^{(1)}$, $R^{(2)}=2j/M^{(2)}$, and $R^{(3)}=2j/M^{(3)}=0.5$ (cf. Tab.~\ref{sumtab}).
The second term corresponds to the interaction and has a model-dependent form
\begin{eqnarray}
{\cal H}'^{(1)}&=&\sqrt{2R^{(1)}+(p^2+x^2)}\,(xy+pq)/\sqrt{2}\,,\nonumber\\
{\cal H}'^{(2)}&=&\sqrt{2R^{(2)}-(p^2+x^2)}\,(xy+pq)/\sqrt{2}\,,\label{h'n}\\
{\cal H}'^{(3)}&=&\sqrt{2R^{(3)}-(p^2+x^2)}\,\sqrt{2}\,xy\,.\nonumber
\end{eqnarray}

The constraints (\ref{em}) and (\ref{emm}) on the conservation of $M^{(1)}$ and $M^{(2)}$ read as
\begin{equation}
p^2+q^2+x^2+y^2=1
\,,
\label{kruh}
\end{equation}
which makes it possible to completely eliminate one degree of freedom in both integrable models.
To do this, we set one of the momenta to zero, in our case $q=0$, and use Eq.~(\ref{kruh}) to fix the corresponding coordinate: $y=\pm\sqrt{1-p^2-x^2}$.
On the level of coherent states, the choice $q=0$ is achieved by considering only a relative phase between $b$ and $c$ bosons, setting the overall phase factor to unity.
Note that this choice is dynamically consistent since the elimination of $y$ ensures that $\dot{q}=\partial{\cal H}/\partial y=0$.

\subsection{Ground-state phase transitions}
\label{qpt}

To analyze the $\aleph\to\infty$ properties of the ground state of the above three models, we set both momenta $p$ and $q$ in Eqs.~(\ref{h0n}) and (\ref{h'n}) to zero, yielding a potential ${\cal V}^{(n)}={\cal H}^{(n)}|_{p=q=0}$ ($n=1,2,3$).
Indeed, any increase of $p$ or $q$ takes one away from the minimum of the Hamiltonian ${\cal H}^{(n)}(p,q,x,y)$ and therefore corresponds to an excitation of the system above the ground state.
The energy 
obtained by the minimization of the potential ${\cal V}^{(n)}(x,y)$ represents an estimate of the scaled ground-state energy ${\cal E}_0=E_0/\aleph$.

The problem can be further simplified for both integrable systems, i.e., for Hamiltonians $H^{(1)}$ and $H^{(2)}$, where the constraint (\ref{kruh}) with $p=q=0$ restricts the ground-state solution to the unit circle $x^2+y^2=1$.
In the SU(1,1) case ($n=1$) with $\omega_0>\omega$, we take
\begin{equation}
x=\sin\vartheta\,,\quad y=\cos\vartheta\,,
\end{equation}
yielding the potential
\begin{equation}
{\cal V}^{(1)}={\cal V}^{(1)}_0+\frac{\Delta\omega}{2}\sin^2\vartheta+\frac{\lambda}{\sqrt{2^3}}\sin(2\vartheta)\sqrt{2R^{(1)}+\sin^2\vartheta}
.
\label{v1}
\end{equation}
For the integrable SU(2) Hamiltonian ($n=2$) with $\omega>\omega_0$, we redefine the angle $\vartheta$ so that
\begin{equation}
x=\cos\vartheta\,,\quad y=\sin\vartheta\,,
\end{equation}
and
\begin{equation}
{\cal V}^{(2)}={\cal V}^{(2)}_0+\frac{\Delta\omega}{2}\sin^2\vartheta+\frac{\lambda}{\sqrt{2^3}}\sin(2\vartheta)\sqrt{2R^{(2)}-\cos^2\vartheta}
.
\label{v2}
\end{equation}
In both cases $\Delta\omega\equiv|\omega_0-\omega|>0$, while ${\cal V}^{(1)}_0=(\omega+R^{(1)}\omega_0)/2$ and ${\cal V}^{(2)}_0=(1-R^{(2)})\omega_0/2$.

\begin{figure}
\includegraphics[width=\linewidth]{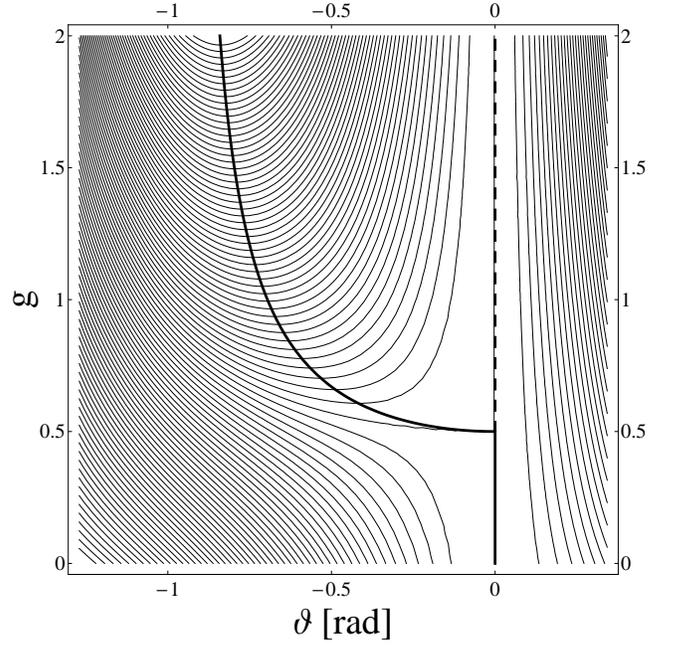}
\caption{The potential energy ${\cal V}^{(1)}(\vartheta)$ and ${\cal V}^{(2)}(\vartheta)$ of the integrable SU(1,1) and SU(2) models as a function of a rescaled coupling parameter $g=\lambda/(\sqrt{2}\Delta\omega)$. The thick curve demarcates the trajectory of the potential minimum, the dashed line indicates the saddle point position.}
\label{f_surf1d}
\end{figure}

For $n=3$, the nonintegrable SU(2) Hamiltonian $H^{(3)}$ of the Dicke model with $\omega_0=\omega$, the constraint (\ref{kruh}) is not applicable.
We take polar coordinates
\begin{equation}
x=r\cos\vartheta\,,\quad y=r\sin\vartheta\,,
\end{equation}
arriving at
\begin{equation}
{\cal V}^{(3)}
={\cal V}^{(3)}_0+\frac{\omega}{2}\,r^2+\frac{\lambda}{\sqrt{2}}\,r^3\sin(2\vartheta)\sqrt{2R^{(3)}-\cos^2\vartheta}
,
\label{v3}
\end{equation}
with ${\cal V}^{(3)}_0=-R^{(3)}\omega/2$.

It is clear that the $M^{(1)}\to\infty$ limit of the SU(1,1) model with $k=\frac{1}{4},\frac{3}{4}$ gives $R^{(1)}\to 0$.
On the other hand, in the integrable version of the SU(2) model it is natural to take $M^{(2)}=4j$, that is $R^{(2)}=0.5$.
Indeed, with this choice the total number of bosons $N$ can be arbitrarily partitioned into $N_b$ and $N_c$, including the extremal choice $(N_b,N_c)=(0,N)$, which corresponds to the photon vacuum combined with a fully excited array of atoms.
With these settings, both expressions (\ref{v1}) and (\ref{v2}) become identical (except the additive constants).
The potential energy surface ${\cal V}^{(1)}$ alias ${\cal V}^{(2)}$ is shown in Fig.~\ref{f_surf1d}.

The minimum of both potentials ${\cal V}^{(1)}$ and ${\cal V}^{(2)}$ for $\lambda=0$ is at $\vartheta=\vartheta_0(0)=0$ (we may equivalently choose $\vartheta_0=\pi$, which would have no influence on the conclusions below).
For increasing $\lambda$, the minimum $\vartheta_0(\lambda)$ remains at the same place until the critical value
\begin{equation}
\lambda^{(1)}_{c0}=\lambda^{(2)}_{c0}=\frac{\Delta\omega}{\sqrt{2}}
\label{cril12}
\end{equation}
is reached.
Here, $\vartheta=0$ becomes a saddle point and the minimum $\vartheta_0(\lambda)$ deviates to negative values, following a trajectory
\begin{equation}
\sin^2\vartheta_0=\frac{12g^2-1-\sqrt{12g^2+1}}{18g^2}
\,,\ 
g\equiv\frac{\lambda}{\sqrt{2}\Delta\omega}\geq\frac{1}{2}
\,.
\label{traj}
\end{equation}
This nonanalytic evolution represents a second-order quantum phase transition.
From Eq.~(\ref{traj}) for $g>g_{c0}=0.5$ we get
$\vartheta_0(g)\sim\sqrt{g-g_{c0}}$ which means that the critical
exponent for the order parameter $\vartheta_0$ (or
$x_0=\sin\vartheta_0$) is equal to $\frac{1}{2}$. 

\begin{figure}
\includegraphics[width=\linewidth]{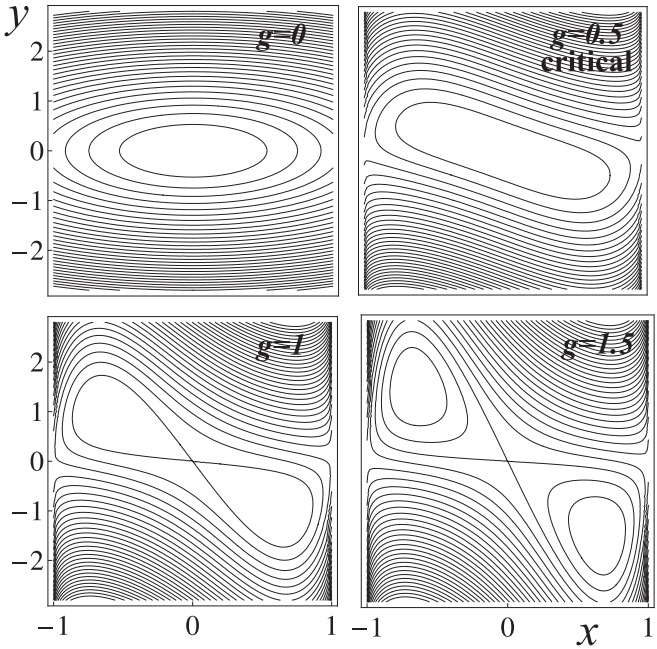}
\caption{The potential energy surface ${\cal V}^{(3)}(x,y)$ of the Dicke model with $\omega_0=\omega$. Values of a rescaled coupling strength $g\equiv\lambda/(\sqrt{2}\omega)$ are given in each panel.}
\label{f_surf2d}
\end{figure}

For the Dicke Hamiltonian $H^{(3)}$ the dimension of the system is not reduced, so the properties of potential (\ref{v3}) must be analyzed in the plane $(r,\vartheta)\equiv(x,y)$.
Nevertheless, the physics is similar to that described above.
For $\lambda$ growing from 0 to a critical value
\begin{equation}
\lambda^{(3)}_{c0}=\sqrt{\frac{\omega\omega_0}{2}}
\label{cril3}
\end{equation}
($\lambda^{(3)}_{c0}=\omega/\sqrt{2}$ for $\omega_0=\omega$) the potential minimum is located at $(x,y)=(0,0)$, which corresponds to a separable state of unexcited atoms and the field vacuum.
At the critical point (\ref{cril3}), the determinant of the Hessian matrix (composed from second derivatives of ${\cal V}^{(3)}$ with respect to both variables) evaluated in the minimum becomes negative, which means that $(x,y)=(0,0)$ becomes a saddle point of the potential.
Starting at this point, two degenerate minima deviate symmetrically to the quadrants with $xy<0$ (for $\lambda>0$).
These minima correspond to a (nearly) degenerate parity doublet of the ground state solutions involving excitations of both atomic and field subsystems.
The distance $r_0$ of the minima from the origin increases with $g\equiv\lambda/\sqrt{2\omega\omega_0}$ as $\sqrt{g-g_{c0}}$ above the critical point $g_{c0}=0.5$, so we have again a second-order QPT with the order parameter $r_0$ characterized by the critical exponent $\frac{1}{2}$.
Various stages of evolution of the potential ${\cal V}^{(3)}$ are shown in Fig.~\ref{f_surf2d}.

\subsection{Excited-state phase transitions}
\label{eqpt}

Any excited-state phase transition can be recognized in the dependence of a quantum level density $\rho({\cal E},\lambda)$ on the scaled energy ${\cal E}$.
At the ESQPT point ${\cal E}_c(\lambda)$ this dependence shows a nonanalyticity whose type enables one to classify the critical behavior in agreement with the standard typology of thermal phase transitions.
The nonanalyticity of $\rho({\cal E},\lambda)$ is reflected by specific discontinuous features in the flow of scaled energy levels ${\cal E}_i(\lambda)$ through the boundary ${\cal E}_c(\lambda)$ \cite{Cap08,Cej08}.

The semiclassical theory of the level density \cite{Gut71} leads to the decomposition
\begin{equation}
\rho({\cal E},\lambda)={\bar\rho}({\cal E},\lambda)+{\tilde\rho}({\cal E},\lambda)
\,,
\label{rhosum}
\end{equation}
where ${\bar\rho}$ and ${\tilde\rho}$ represent smooth and oscillatory components, respectively.
The oscillatory component can be expressed as a sum over periodic orbits in the general form ${\tilde\rho}=\sum_kA_k\cos(\hbar^{-1}S_k+\phi_k)$, where $A_k$ and $\phi_k$ stand for an amplitude and a phase shift of the $k$th orbit contribution, while $S_k=\oint\vec{p}\cdot d\vec{x}$ represents the action over this orbit.
In the limit $\hbar\to 0$ this part leads to infinitely rapid oscillations which cancel out if the level density is integrated over an arbitrary narrow interval of energy.
In this limit, only the smooth component of Eq.~(\ref{rhosum}) is relevant.
It is expressed via orbits of zero length, yielding the formula
\begin{equation}
{\bar\rho}({\cal E},\lambda)=(2\pi\hbar)^{-f}\underbrace{\int\delta\bigl({\cal E}-{\cal H}(\vec{p},\vec{x},\lambda)\bigr)d\vec{p}\,d\vec{x}}_{\Omega({\cal E},\lambda)}
\,,
\label{rhovol}
\end{equation}
where ${\cal H}$ is the classical Hamiltonian depending in general on $f$-dimensional vectors of coordinates $\vec{x}$ and momenta $\vec{p}$.
The quantity $\Omega({\cal E},\lambda)\,d{\cal E}$ represents a $2f$-dimensional volume of the available phase space for the interval of energy $({\cal E},{\cal E}\!+d{\cal E})$.

As follows from these considerations, in the systems with synonymous thermodynamic ($\aleph\to\infty$) and classical ($\hbar\to 0$) limits any kind of nonanalyticity in the energy dependence of the classical phase space volume generates an ESQPT on the quantum level.
Such nonanalyticities may follow, e.g., from the presence of the Hamiltonian stationary points \cite{Cej06,Cap08,Cej08,Ley05,Rei05,Car92,Hei02,Bab09}.
Because the quantum microcanonical entropy is proportional to a logarithm of $\rho({\cal E},\lambda)$, the dependence of the level density on the scaled energy represents a key for the ESQPT classification (consistent with the classification of the corresponding thermal phase transition).
For instance, a jump of $\Omega({\cal E},\lambda)$ at ${\cal E}_c(\lambda)$ corresponds to a first-order phase transition.
It is known that in systems with $f=1$ there exist ESQPT effects even stronger than those of the first-order type.
These effects are associated with an infinite peak of $\Omega({\cal E},\lambda)$, which shows up the corresponding dependence $\rho({\cal E},\lambda)$.
The origin of this behavior is often found in a local maximum of the one-dimensional potential at energy ${\cal E}_c(\lambda)$.
On the other hand, a softer type of nonanalyticity in $\Omega({\cal E},\lambda)$, like a discontinuous or infinite derivative, causes a continuous phase transition.
This typically happens in the systems with more than one degrees of freedom \cite{Cej08}.
In case of a discontinuous $(n-1)$th derivative ($n\geq 2$) the transition is of $n$th order, and for a singular (infinite) derivative the transition has no Ehrenfest classification.

\begin{figure}
\includegraphics[width=\linewidth]{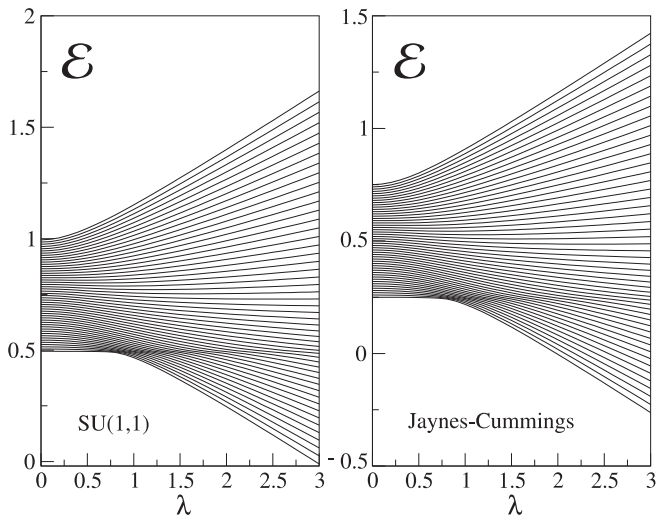}
\caption{Level dynamics for the SU(1,1) model (left) and for the SU(2) integrable model (right) with $M^{(1)}=M^{(2)}=100$ and $\Delta\omega=1$. The scaled energies were obtained by an exact diagonalization. The ESQPT above the ground state critical point $\lambda_{c0}=0.707$ is apparent in the bunching of levels around critical energies ${\cal E}^{(1)}_c=0.5$ and ${\cal E}^{(2)}_c=0.25$, respectively.} 
\label{f_ldyn1}
\end{figure}

The ESQPT effects are present in the spectra of all three models described above.
We saw in Sec.~\ref{qpt} that the ground-state QPTs are located at the critical points $\lambda^{(n)}_{c0}$ from Eqs.~(\ref{cril12}) and (\ref{cril3}).
It turns out that for $\lambda>\lambda^{(n)}_{c0}$ the singularity propagates into the excited spectrum.
Let us first consider the integrable Hamiltonians $H^{(1)}$ and $H^{(2)}$, both corresponding to an effectively one-dimensional configuration space.
When the critical point is reached in these systems, the global minimum of the potential ${\cal V}^{(1)}$ or ${\cal V}^{(2)}$ changes into a saddle point, which remains present for all values $\lambda>\lambda^{(1)}_{0c}$ or $\lambda^{(2)}_{0c}$.
The saddle point represents a singularity of $\Omega({\cal E},\lambda)$, causing the strongest type of ESQPT characterized by the infinite peak in the semiclassical level density.
To see this, recall that in systems with $f=1$ the integral in Eq.~(\ref{rhovol}) is equal to the period $\tau$ of the single (uniquely determined) classical orbit at energy ${\cal E}$, hence $\Omega({\cal E},\lambda)=\tau({\cal E},\lambda)$.
If ${\cal E}$ coincides with the energy of the $x=0$ ($\vartheta=0$) saddle point,
the period becomes infinite
because $x=p=0$ is a stationary point ($\dot{x}=\dot{p}=0$) of both Hamiltonians ${\cal H}^{(1)}(p,x)$ and ${\cal H}^{(2)}(p,x)$.
Let us note that the same type of ESQPT is observed in systems with one quantum degree of freedom showing a local maximum of the potential, for instance in the Lipkin model \cite{Ley05,Rib07,Rel08} and many others \cite{Cap08,Cej08}.

\begin{figure}
\includegraphics[width=\linewidth]{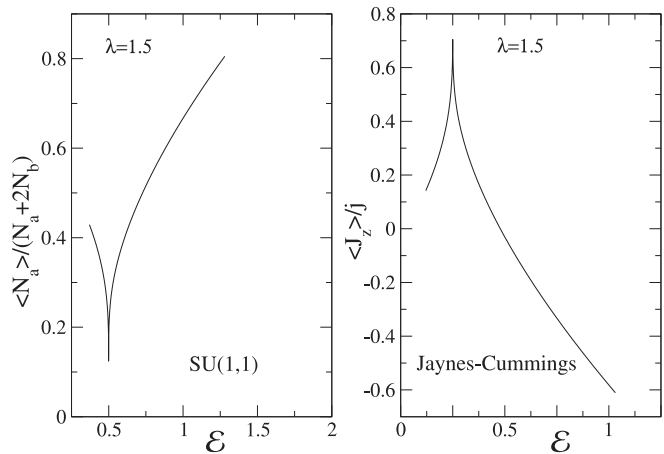}
\caption{Expectation values of $N_a$ (left) and $J_z$ (right) for individual states across the spectrum at $\lambda=1.5$. The left and right panels, respectively, correspond to the SU(1,1) and the SU(2) integrable models with $M^{(1)}=M^{(2)}=2000$. The ESQPT is indicated by needlelike singularities located at the critical energies.}
\label{f_esqpt1d}
\end{figure}

\begin{figure*}
\includegraphics[width=0.8\linewidth]{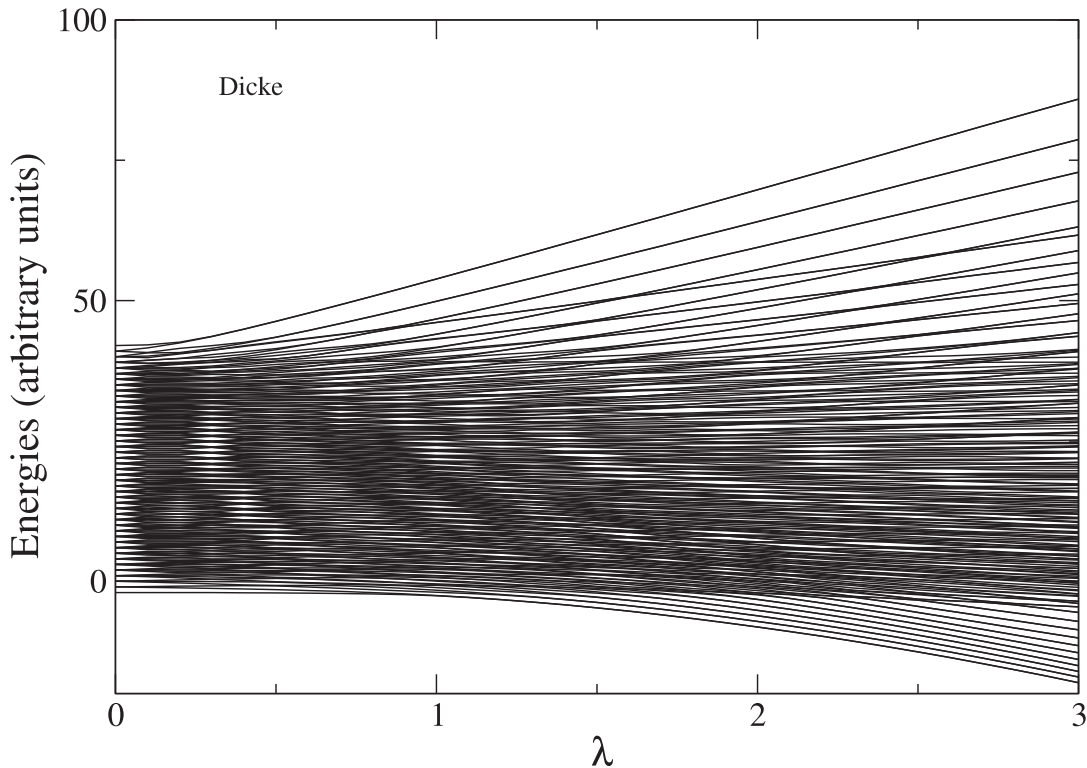}
\caption{Level dynamics for the SU(2) nonintegrable model $H^{(3)}$ with $j=2$, obtained by a numerical diagonalization with $N_{\rm trunc}\approx 40$. We show absolute energies in units of $\omega=\omega_0$. A steep growth of the level density at the energy $E=-2$ [corresponding to the saddle-point of potential (\ref{v3})] indicates a continuous ESQPT in the infinite-size limit.}
\label{f_ldyn2}
\end{figure*}

A local increase of the level density at the saddle-point energy ${\cal E}^{(n)}_c={\cal V}^{(n)}_0$ in both integrable models, i.e., the SU(1,1) and Jaynes-Cummings models ($n=1$ and 2, respectively), is demonstrated in Fig.~\ref{f_ldyn1}.
The two panels capture the evolution of quantum spectra for both models with the interaction parameter $\lambda$, showing clear indications of the ground-state QPT and its extension into the ESQPT on the right-hand side of the critical point, which for $\Delta\omega=1$ is at $\lambda^{(1)}_{c0}=\lambda^{(2)}_{c0}=0.707$ (see Tab.~\ref{sumtab}).
The calculation was done in a finite-size case, but it shows well pronounced precursors of the phase transitional behavior.

Additional ESQPT signatures are depicted in Fig.~\ref{f_esqpt1d}, which shows expectation values of operators proportional to
$K_0=N_c+k$ and $J_0=N_c-j$ in individual excited states as a function of scaled energy ${\cal E}$ for a fixed value of $\lambda=1.5$.
Specifically, we consider the operator $N_a/M^{(1)}=(2K_0-\frac{1}{2})/M^{(1)}$ for the SU(1,1) model and $J_0/j$ for the Jaynes-Cummings model.  
Note that these two observables act as order parameters of the respective standard QPTs: their ground-state expectation values change from $\ave{N_a}_0=0$ to $\ave{N_a}_0>0$, and from $\ave{J_0}_0=j$ to $\ave{J_0}_0<j$, as $\lambda$ crosses the critical point $\lambda_{c0}$.

In both panels of Fig.~\ref{f_esqpt1d}, the respective system is well above the QPT critical point.
The energy dependence of the respective expectation value $\ave{N_a}_{\cal E}$ and $\ave{J_0}_{\cal E}$ shows a cusplike shape with a singularity localized at the ESQPT energy ${\cal E}_c^{(n)}$ (cf. Fig.~\ref{f_ldyn1}).
The two shapes are mutually reversed: while for the SU(1,1) model, the expectation value drops sharply to the lowest value at the critical energy, for the SU(2) model it has a needle-shaped maximum.
This is connected with a singular localization of the semiclassical wave function for ${\cal E}={\cal E}^{(n)}_c$ at the saddle point of the potential, i.e., at $x=0$ for the SU(1,1) and $x=1$ for SU(2) model (in both cases $\vartheta=0$).
This implies $\ave{N_c}_{{\cal E}_c}=0$ for the SU(1,1) case (hence $\ave{N_a}_{{\cal E}_c}=0$ or 1 for even or odd systems, respectively) and $\ave{N_c}_{{\cal E}_c}=\frac{1}{2}M^{(2)}$ for the SU(2) case (so $\ave{J_0}_{{\cal E}_c}=\frac{1}{2}M^{(2)}-j$).
An analogous effect (explained by infinite dwell times of a classical particle at the stationary point) is known from one-dimensional systems with a local maximum of the potential \cite{Car92}.
The ESQPT critical energies ${\cal E}^{(1)}_c$ and ${\cal E}^{(2)}_c$ drop to the ground-state energy as $\lambda$ decreases to the respective critical points $\lambda_{c0}^{(1)}$ and $\lambda_{c0}^{(2)}$, and so do both cusp singularities in Fig.~\ref{f_esqpt1d}.
Below the critical point the singularities disappear.

For the nonintegrable Dicke model with the Hamiltonian $H^{(3)}$, the two-dimensional potential (\ref{v3}) has a saddle point at $(x,y)=(0,0)$.
This is connected with a nonanalytic dependence of the phase space volume (and the level density) on the scaled energy, although of a softer type than in the previous case, as follows from a higher dimensionality of the phase space for the Dicke model.
Specifically, for $\lambda>\lambda^{(3)}_{c0}$ the level density exhibits an anomalous growth with an infinite derivative (singular tangent) at ${\cal E}={\cal E}^{(3)}_c$, which coincides with the saddle-point energy ${\cal V}^{(3)}_c$ of the potential \cite{Cej08}.
The resulting ESQPT is continuous (but without the Ehrenfest classification), although its finite-size precursors very much resemble those of a first-order phase transition (the level density is close to a step-like function).

The step-like increase of the level density in the Dicke model can be seen in Fig.~\ref{f_ldyn2}, where the level dynamics with variable $\lambda$ is shown for $j=2$.
Even for such a moderate value of the angular momentum, a sharp precursor of the ESQPT effect at absolute energy $E_c^{(3)}=M^{(3)}{\cal E}_c^{(3)}=-1$ is well visible in the spectrum above $\lambda^{(3)}_{c0}=0.707$ (for $\omega=\omega_0=1$) as the lower interface between the horizontal and sloped level contours.
Note that the effects of the Hilbert space truncation (the cutoff for the number of photons; see Sec.~\ref{nume}) become relevant for the high-energy part of the spectrum.

\section{Quench dynamics}
\label{dyna}

\subsection{Survival probability and energy distribution}
\label{sura}

The Hamiltonians introduced in Sec.~\ref{mod} have the common form (\ref{hhh}), that is ${\cal H}(\lambda)={\cal H}_0+\lambda\,{\cal H}'$ if omitting the model specifying superscript $n$.
Here ${\cal H}_0$ and ${\cal H}'$ represent the free and interaction term, respectively, and $\lambda$ is a dimensionless control parameter.
Let us stress that here we are working with the scaled Hamiltonian ${\cal H}=H/\aleph$, but consider a finite-$\aleph$ case, so that in general $[{\cal H}_0,{\cal H}']\neq 0$.
As seen from the expression ${\cal H}(\lambda_2)={\cal H}(\lambda_1)+\Delta\,{\cal H}'$ with $\Delta=\lambda_2-\lambda_1$, the above Hamiltonian allows one to apply perturbation techniques with the same perturbation ${\cal H}'$ for all initial points $\lambda_1$.

Suppose that the system is initially prepared in one of the eigenstates $\ket{\psi_i(\lambda_1)}\equiv\ket{\psi_1}$ of ${\cal H}(\lambda_1)\equiv {\cal H}_1$ with energy $E_1(\lambda_1)/\aleph\equiv{\cal E}_1$.
Below we will consider the initial state $\ket{\psi_1}$ coinciding with the ground state $\ket{\psi_0(\lambda_1)}$, but the formalism can be very easily developed for the general case.
At time $t=0$, the value of the control parameter is abruptly changed from $\lambda_1$ to $\lambda_2=\lambda_1+\Delta$.
The state $\ket{\psi_1}$ is no more an eigenstate of the new Hamiltonian ${\cal H}(\lambda_2)\equiv {\cal H}_2$ and starts evolving.

The evolution after the quench can be monitored by a survival probability $p_1(t)=|a_1(t)|^2$, where
\begin{equation}
a_1(t)=\matr{\psi_1}{e^{-i{\cal H}_2t}}{\psi_1}=\int\underbrace{\left|\scal{{\cal E}_2}{\psi_1}\right|^2}_{\omega_1({\cal E}_2)}e^{-i{\cal E}_2t}d{\cal E}_2
\label{surv}
\end{equation}
is an amplitude describing the decay and recurrence of the initial state $\ket{\psi_1}$ for $t>0$.
A formula of this form captures in general all quantum decay processes and has been studied in many different contexts (e.g., in analyses of the fidelity or Loschmidt echo \cite{Pro}).
Note that the use of the scaled Hamiltonian ${\cal H}_2$ in Eq.~(\ref{surv}) is equivalent to the $t\to t/\aleph$ transformation of time in the expression with unscaled Hamiltonian $H_2$.
Expanding the initial state $\ket{\psi_{1}}$ in the eigenbasis $\ket{{\cal E}_{2i}}\equiv\ket{{\cal E}_{i}(\lambda_2)}$ of the Hamiltonian ${\cal H}_2$ (with $i=1,2,\dots$ enumerating discrete eigenvalues ${\cal E}_{2i}$),
\begin{equation}
\ket{\psi_{1}}=\sum_{i}\underbrace{\scal{{\cal E}_{2i}}{\psi_1}}_{c_i} \ket{{\cal E}_{2i}} 
\,,
\label{ces}
\end{equation}
the survival probability reads as
\begin{equation}
p_1(t)=\sum_i|c_i|^4+2\sum_{i>j}|c_i|^2|c_j|^2\cos[({\cal E}_{2i}-{\cal E}_{2j})t]
\,.
\label{surv2}
\end{equation}

As indicated in Eq.~(\ref{surv}), the survival amplitude $a_1(t)$ can be written as the Fourier transform of the energy distribution $\omega_1({\cal E}_2)\equiv|\scal{{\cal E}_2}{\psi_1}|^2$ of the initial state in the eigenbasis of ${\cal H}_2$.
The precise energy distribution is given by
\begin{equation}
\omega_1({\cal E}_2)\equiv\sum_i|c_i|^2\delta({\cal E}_2-{\cal E}_{2i})
\,.\label{four}
\end{equation}
Since both functions $p_1(t)$ in Eq.~(\ref{surv2}) and $\omega_1({\cal E}_2)$ in Eq.~(\ref{four}) are expressed in terms of the discrete energies ${\cal E}_{2i}$ and the corresponding occupation probabilities $|c_i|^2$, they comprise fully equivalent information on the quench-induced relaxation process.

The discrete form (\ref{four}) of the energy distribution $\omega_1({\cal  E}_2)$ can be approximated by its smoothened form ${\bar\omega}_1({\cal  E}_2)$, obtained by replacing the $\delta$ functions by normalized Gaussian profiles centered at eigenenergies ${\cal E}_{2i}$.
This leads to  
\begin{equation}
{\bar\omega}_1({\cal  E}_2)=\sum_{i}|c_i|^2\frac{1}{\sqrt{2\pi\sigma_i^2}}\exp\left[-\frac{({\cal E}_2-{\cal E}_{2i})^2}{2\sigma_i^2}\right]
,
\label{gauss}
\end{equation}
where the widths $\sigma_i$ are chosen separately for each Gaussian with regard to the local density of states in the respective part of the spectrum of ${\cal H}_2$.
The aim of the smoothening procedure is to overcome the discrete character of $\omega_1({\cal  E}_2)$ while loosing as little as possible information on its local behavior.
We therefore set the width of each Gaussian to the spacing between the ($i$+1)th and $i$th levels, so $\sigma_{i}={\cal E}_{2(i+1)}-{\cal E}_{2i}$.

Although the discrete and smoothed forms (\ref{four}) and (\ref{gauss}) capture basically the same information, their distinction leads to two visualization methods of the energy distribution.
In the first one, based on the discrete form $\omega_1({\cal  E}_2)$, the values of $|c_i|^2$ are drawn against ${\cal E}_{2i}$ in the form of a scatter plot (individual points being enumerated by the eigenvalue index $i$).
The other method shows the smoothened distribution ${\bar\omega}_1({\cal  E}_2)$ as a continuous function of energy ${\cal E}_2$.
While the first method displays essentially the average of $|c_i|^2$ in the given energy domain irrespective of the number of states (level density) in this domain, the second method inherently contains a density-dependent weighting.
In the following, we use both methods and compare the resulting forms with the time evolution of the survival probability $p_1(t)$, calculated from the exact formula (\ref{surv2}).

\subsection{A critical quench}
\label{cri}

Rather specific shapes of the energy distributions $\omega_1({\cal E}_2)$ and ${\bar\omega}_1({\cal  E}_2)$ can be expected if the system exhibits an excited-state quantum phase transition for $\aleph\to\infty$.
Assume that a sequence of such transitions is indeed present at energies ${\cal E}_c(\lambda)$ depending, in general, on the control parameter $\lambda$ (we know that in the models studied here, ${\cal E}_c$ is a constant).
The critical curve ${\cal E}_c(\lambda)$ in the plane ${\cal E}\times\lambda$ may eventually reach the lowest energy of the system; then the ground-state quantum phase transition is observed at the corresponding value $\lambda=\lambda_{c0}$ (it is so in the present models).
As discussed above, the flow of energy levels as a function of $\lambda$ and the density of the spectrum as a function of ${\cal E}$ are nonanalytic when crossing the ESQPT critical curve.
Therefore, if the energy distribution $\omega_1({\cal E}_2)$ or ${\bar\omega}_1({\cal  E}_2)$ of the initial state after the quantum quench interferes with the critical value ${\cal E}_c(\lambda_2)$, one may expect some anomalous properties of the survival probability.

We can easily estimate which parameter changes $\Delta=\lambda_2-\lambda_1$ may lead to such anomalous relaxation processes.
To do so, recall that the mean value $\overline{{\cal E}}_2$ of both energy distributions $\omega_1({\cal E}_2)$ and ${\bar\omega}_1({\cal  E}_2)$ (both forms yield the same value) is given by $\overline{{\cal E}}_2=\matr{\psi_1}{{\cal H}_2}{\psi_1}$, so
\begin{equation}
\overline{{\cal E}}_2={\cal E}_1+\Delta\underbrace{\matr{\psi_1}{{\cal H}'}{\psi_1}}_{{\cal E}'_1}\,.
\label{av}
\end{equation}
The \uuvo{critical quench} $\Delta_c$ for a given initial state is the one for which the average $\overline{{\cal E}}_2$ coincides with the critical value ${\cal E}_c(\lambda_2)$, hence
\begin{equation}
\Delta_c=\frac{{\cal E}_c(\lambda_1+\Delta_c)-{\cal E}_1}{{\cal E}'_1}\,.
\label{cq}
\end{equation}
Of course, an actual range of the ESQPT-influenced quenches covers a wider interval of the $\Delta$ values around $\Delta_c$, depending on the width of the distribution $\omega_1({\cal E}_2)$ or ${\bar\omega}_1({\cal  E}_2)$ and also on the smearing effects in the phase-transitional signatures due to the actual finite value of the size parameter $\aleph$.
Nevertheless, the above formula yields a good estimate of a central point of the interval, where the quench dynamics can be expected to show strong ESQPT precursors.

\begin{figure}
\includegraphics[width=0.7\linewidth,angle=-90]{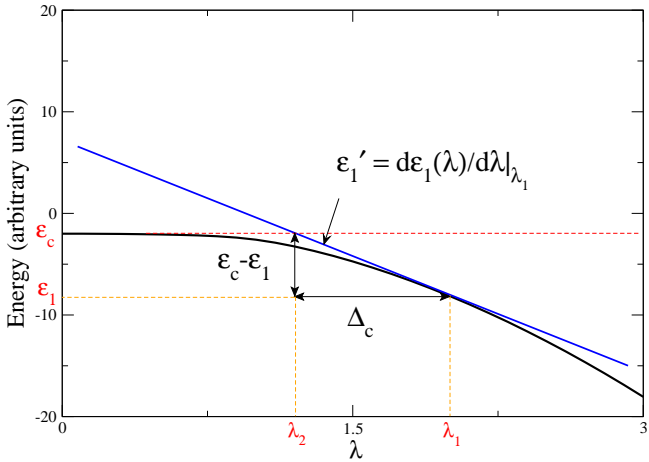}
\caption{Graphical determination of the critical quench $\lambda_1\to\lambda_2$ for a given initial state: the value of $\Delta_c$ is given by an intersection of the tangent ${\cal E}'_1$ with the critical \uuvo{curve} ${\cal E}_c$.}
\label{crqu}
\end{figure}

The application of the Hellman-Feynman theorem to Eq.~(\ref{av}) yields $\matr{\psi_1}{{\cal H}'}{\psi_1}\equiv{\cal E}'_1=d{\cal E}_i(\lambda)/d\lambda|_{\lambda=\lambda_1}$.
This leads to a simple graphical interpretation of Eq.~(\ref{cq}) shown in Fig.~\ref{crqu}. 
The parameter change $\lambda_1\to\lambda_2$ is identified with the critical quench if the tangent of the initial energy level ${\cal E}_i(\lambda)$ at $\lambda=\lambda_1$ crosses the critical curve ${\cal E}_c(\lambda)$ at $\lambda=\lambda_2$.
Let us stress that the final value of the control parameter corresponding to the critical quench, $\lambda_2=\lambda_1+\Delta_c$, differs in general from the critical value $\lambda_{c0}$ of the ground-state quantum phase transition.

\subsection{Results for integrable models}
\label{rein}

Now we are ready to discuss model-specific results for the energy distributions (\ref{four}) and (\ref{gauss}) and the corresponding survival probability based on Eq.~(\ref{surv}).
Note that in this and in the following section we return to the unscaled energy $E=M^{(n)}{\cal E}$.

\begin{figure*}
\includegraphics[angle=0,width=0.7\linewidth]{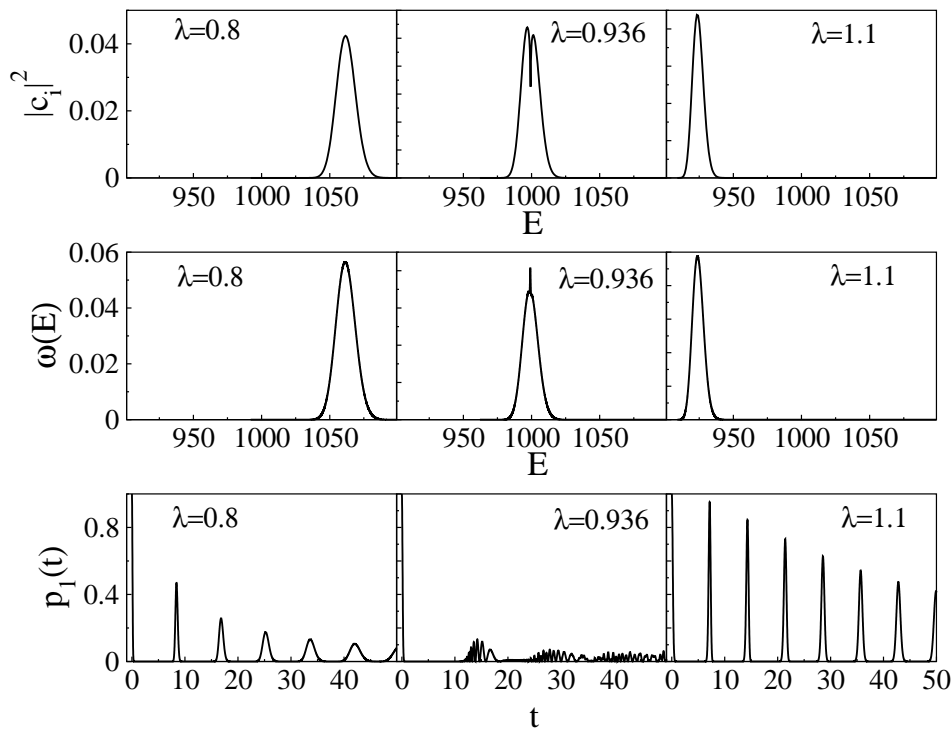}
\caption{
Energy distributions and survival probabilities for three quantum quenches in the SU(1,1) model ($M^{(1)}=2000$). The initial state is the ground state at $\lambda_1=1.5$ and the final parameter values $\lambda_2$ are given in each panel. The leftmost and rightmost panels in each row correspond to quenches above and below the critical energy, respectively, while the middle panel depicts a quench to the critical region. Upper row: the energy distribution of probabilities $|c_i|^2$, see Eq.~(\ref{four}). Middle row: the smoothened energy distribution from Eq.~(\ref{gauss}). Lower row: the survival probability from Eq.~(\ref{surv}).
}
\label{f_ressu1}
\end{figure*}

We start with the two integrable models.
Figure~\ref{f_ressu1} shows results for three quenches in the SU(1,1) model with $2N_{b}+N_{a}=2000$ (thus $N_a$ even).
The initial state is identified with the ground state at $\lambda=\lambda_1=1.5$ and the respective final parameter value $\lambda_2$ is written separately in each panel.
In the upper row of panels we present the quantity $\omega_1(E_2)$ as a scatter plot of the values $|c_i|^2$ versus the energy eigenvalue $E_{2i}$.
Note that the number of points is so large here that the scatter plots look like continuous curves.
The panels from left to right correspond to a quench above, at, and below the critical energy, which for the present setting coincides with $E_c=1000$.
While for both noncritical quenches (left and right panels) the distribution of $|c_i|^2$ exhibits just a single peak centered at energy ${\overline E}_2$ depending on the value of $\lambda_2$, the critical quench to the final value $\lambda_2=0.936$ (middle panel) leads to a more complex distribution.
In this case we observe a double peak structure in the plot of $|c_i|^2$, the peak-separating minimum being localized exactly at the ESQPT energy.

The criticality of the quench to $\lambda_2=0.936$ can also be seen in the other rows of panels in Fig.~\ref{f_ressu1}.
In the second row, we show the smoothened distribution ${\bar\omega}_1(E_2)$ from Eq.~(\ref{gauss}).
A clear difference from the first row is observed in the middle panel, where the second row shows a sharp maximum at the critical energy, in contrast the minimum in the first row.
This is due to the above-discussed (Sec.~\ref{sura}) distinction between the visualization methods based on the raw and smoothened energy distributions $\omega_1(E_2)$ and ${\bar\omega}_1(E_2)$.
We know (Sec.~\ref{eqpt}) that there is a local increase of the level density around the critical energy $E\approx E_c$ (see Fig.~\ref{f_ldyn1}) connected with diverging periods of the classical trajectories passing the saddle point.
This leads to a sizable increase of the distribution ${\bar\omega}_1(E_2)$, despite the fact that individual values of $|c_i|^2$ are lower in the critical region, as seen in the upper panel of Fig.~\ref{f_ressu1}.

In the lower row of panels in Fig.~\ref{f_ressu1} the survival probability $p_1(t)$ is shown as a function of time for the three quenches discussed above.
Again, similar patterns are observed for both noncritical quenches (left and right panels). 
In these cases, the survival probability exhibits regular damped oscillations.
The time constant $\tau$ of the decaying envelope is related to the total width $\Delta E$ of the associated peak in the energy distribution by the Heisenberg-like relation $\tau\propto 1/\Delta E$, while the frequency and form of particular oscillations depend on the mean energy and the fine structure of the energy distribution.
For the critical quench (middle panel), the survival probability behaves differently than for the noncritical cases.
The quick initial decay is followed just by small random oscillations in the region $p_1(t)\approx 0$, avoiding the slowly damped recurrences present in the other panels.
This type of dynamics is connected with the above-discussed modified form of the energy distribution shown in the upper panels of Fig.~\ref{f_ressu1}.

It needs to be stressed that we are dealing here with a finite system whose behavior is unavoidably quasiperiodic.
Hence, strictly speaking, the lack of recurrences seen in the middle low panel of Fig.~\ref{f_ressu1} can only be temporal, as follows from the exact formula for the survival probability in Eq.~(\ref{surv2}).
It is known, however, that in realistic situations the quasiperiodicity of quantum evolution on the long time scales is beaten by decoherence effects \cite{Ber}. 
The short and medium time scales addressed in the present calculations are therefore most substantial from the practical viewpoint.
We interpret the observed difference in the character of the quench-induced relaxation process as an important dynamical consequence of the ESQPT.

Rather similar results are obtained also for the SU(2)-based integrable model.
In Fig.~\ref{f_resjcu}, we show the same quantities as in the previous figure, but for the Jaynes-Cummings model and only for the critical quench.
The calculation was done with $j=500$ ($M^{(2)}=2000$), so the ESQPT critical energy $E_c=500$.
Again, the initial state coincides with the ground state at $\lambda_1=1.5$ and the final parameter value $\lambda_2=0.936$, for which the results are shown, corresponds to the critical case.
The energy distribution $\omega_1(E_2)$ (a scatter plot of $|c_{i}|^2$ values) is given in the upper right panel of Fig.~\ref{f_resjcu}, the smoothened energy distribution $\omega_1(E_2)$ in the upper left panel, and the survival probability $p_{1}(t)$ in the lower panel.
We observe essentially the same behavior as in the middle column of panels in Fig.~\ref{f_ressu1}.
This is not surprising since the two models have a rather similar structure.

In summary, it is clear that the presence of an excited-state phase transition in the spectrum of both integrable models has major impact on the quench-induced relaxation processes. 
We observe that the survival probability quickly decays and shows no recurrences in the medium time scale for the QQs which lead the system to the ESQPT critical energy.
This behavior gives a strong support to the conjecture proposed (in connection with the Lipkin model) in Ref.~\cite{Rel08}.

\subsection{Results for the nonintegrable model}
\label{renoin}

\begin{figure}[t]
\includegraphics[angle=-90,width=\linewidth]{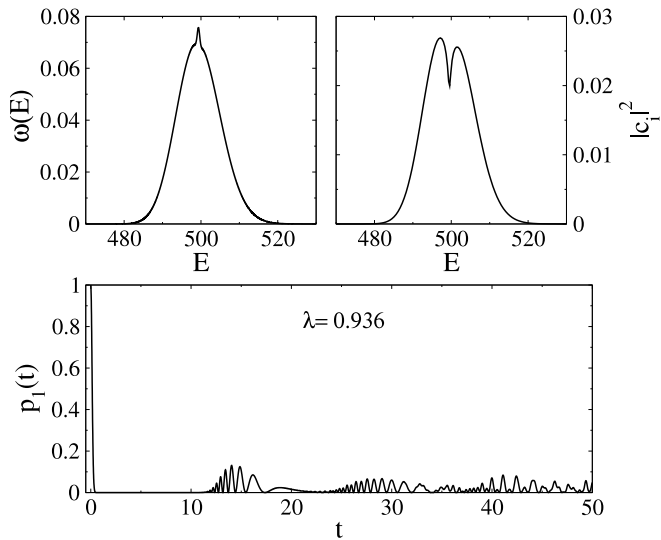}
\caption{The same quantities as in Fig.~\ref{f_ressu1}, but for the critical quench in the Jaynes-Cummings model with $M^{(2)}=2000$.}
\label{f_resjcu}
\end{figure}

\begin{figure}
\includegraphics[angle=-90,width=\linewidth]{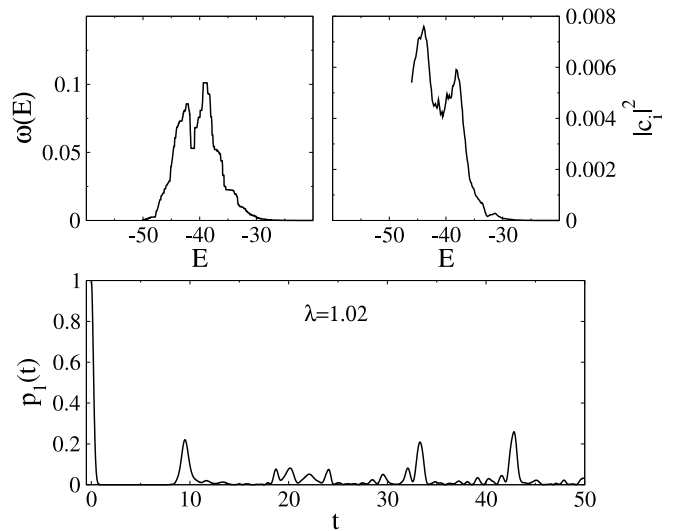}
\caption{The same as in Fig.~\ref{f_resjcu}, but for the nonintegrable Dicke model with $j=40$. The curve in the upper-right panel is cut on the low-energy side because of the additional smoothening procedure.}
\label{f_resdic}
\end{figure}

The nonintegrable Dicke model shows more complex behavior than the two integrable models discussed above, and is also more difficult from the numerical point of view.
We show in Fig.~\ref{f_resdic} the results obtained for the critical quench in this model.
The system is defined by $j=40$ (i.e., it contains 80 atoms) and by the resonance condition $\omega_{0}=\omega=1$.
In this case, the absolute energy corresponding to the ESQPT is $E_{c}=-40$.
As in the previous cases, the initial state before the quench is the ground state at $\lambda_1=1.5$, while the final parameter value corresponding to the critical quench is $\lambda_2=1.02$.
The arrangement of Fig.~\ref{f_resdic} is the same as that of Fig.~\ref{f_resjcu}.

As discussed in Sec.~\ref{nume}, the infinite dimension of the Hilbert space of the Dicke model requires to pay an appropriate attention to the convergence issues.
In our case, the stability of results against the truncation of the Hilbert space was checked by varying the cutoff parameter $N_{\rm trunc}$ for the number of $b$ bosons (photons) until the convergence was reached for the quantities considered. 
In practice, one has to perform several runs of the computation with increasing value of $N_{\rm trunc}$ and compare the results obtained in each run.
It needs to be stressed that an optimal value of the cutoff parameter (satisfying a plausibly defined convergence criterion) depends on the quantities considered and particularly on the relevant range of energy.
The calculations presented in Fig.~\ref{f_resdic} (with 80 atoms) were done including all states of the photon field up to $N_b=N_{\rm trunc}=220$.
The dimensions connected with these high particle numbers are at the limit of our present computing capabilities.
 
Due to the nonintegrability of the Dicke model, the behavior observed in Fig.~\ref{f_resdic} is partly different from the behavior of the same quantities in Fig.~\ref{f_resjcu}.
In particular, the energy distribution of the $|c_{i}|^2$ probabilities and the smoothened distribution $\omega_1(E_2)$ exhibit much stronger fluctuations than those of the integrable models.  
For this reason, an additional smoothing procedure (different from the one described in Sec.~\ref{sura}) has been applied to the result of the calculation. 
The smoothed energy distributions are presented in the upper panels of Fig.~\ref{f_resdic}. 
As we see in the upper left panel, the local maximum of $\omega_1(E_2)$ at the critical energy, clearly observed in both integrable models, is lost. 
However, the main feature of both energy distributions, which is their split form around the ESQPT critical energy, is well reproduced.
  
Concerning the survival probability after the critical quench (the lower panel of Fig.~\ref{f_resdic}), it behaves similarly as in the above integrable models. 
It should be noted, however, that for the integrable models the survival probability yields its characteristic \uuvo{critical shape} (with no recurrences) only in a very narrow interval around the value of $\lambda_2$ defining the critical quench.
In contrast, the Dicke model yields a much wider interval of critical-like relaxation responses.
Only for $\lambda_2$ far away from the critical-quench value (smaller or larger), the $|c_{i}|^2$ and $\omega_1(E_2)$ distributions receive their typical single-peak shapes and the survival probability $p_1(t)$ gets the corresponding form of damped oscillations.

We may conclude that the ESQPT in the Dicke model affects the quench dynamics in a qualitatively similar way as in the integrable models, although its fingerprints in various QQ-related observables are fuzzier than those discussed in Sec.~\ref{rein}.
The observed differences are partly due to the fact that the ESQPT in the Dicke model is of a softer type than those in the SU(1,1) and Jaynes-Cummings models (see Sec.~\ref{eqpt}).
Another reason for the softening is the chaotic nature of dynamics in the Dicke model \cite{Ema03}, which, generically, has a tendency to obscure the ESQPT signatures \cite{Cej08}.

\section{Conclusions}
\label{con}

We have studied the phase diagram and the nonequilibrium dynamics of three models describing the interaction of a single-mode bosonic field with an algebraic subsystem based on either the SU(1,1), or the SU(2) algebras.

The existence of an excited-state quantum phase transition in both integrable SU(1,1) and SU(2)-based quantum models 
is revealed for finite systems as a local peak in the level density, which in the thermodynamic limit transforms into a singularity.
In the nonintegrable SU(2)-based model, the ESQPT leads to a step-like increase of the level density, which limits to a dependence with an infinite derivative.
The ESQPT manifests itself also in the expectation values of quantum observables that depict singularities at the critical scaled energy. 
These signals of the presence of an ESQPT open the possibility of using the concept of order parameter and to resort to the Landau theory to characterize and classify them. 

We have investigated the consequences of an ESQPT on the relaxation dynamics after a quantum quench.
Starting from an initial state that we choose as the ground state of the system for a specific value of the control parameter, a sudden change of the control parameter is applied and the relaxation process is followed by solving the time-dependent Schr{\"o}dinger equation.
This is done either exactly (for the integrable models), or in a truncated space (for the Dicke model), where the convergence issues are taken into account.

Various relevant magnitudes related to the relaxation process after the quench are studied. 
In particular, we analyze the survival probability of the initial state after the quench, which is closely related to the energy distribution of the initial state in the eigenbasis of the new Hamiltonian.
We see a dramatic effect in the survival probability for the {\em critical quench\/} that drives the system from the initial ground state to the critical energy domain associated with the ESQPT at the new value of the control parameter.
This effect is studied separately in the three models used.

The two integrable systems subjected to normal (noncritical) quenches display the typical pattern of collapses and revivals with a smooth decaying envelope, as follows from the single-peak forms of the respective energy distributions.
In contrast, the critical quench produces a sudden destruction of the survival probability followed by small random oscillations. 
This specific response is connected to a more complex shape of the energy distribution, showing a kind of splitting right at the ESQPT energy.

A similar phenomenon is also observed in the nonintegrable Dicke model. 
However, due to level repulsion the survival probability is reduced in amplitude and the critical region is much broader. 
Away from the critical region, the survival probability and the energy distribution behave in a similar way as in the integrable models. 

In spite of the differences observed in integrable and nonintegrable cases, we believe that relaxation dynamics offers clear signals of an excited-state phase transitions in the Dicke type of models.
The character of these transitions deserves more studies.

\section*{Acknowledgements}

This work has been partially supported by the Czech Science Foundation (202/09/0084), by the Czech Ministry of Education (MSM 0021620859), by the Spanish Mi\-nis\-terio de Educaci\'on y Ciencia and the European regional development fund FEDER (FIS2009-07277, FIS2008-04189, FIS2006-12783-C03-01, FPA2006-13807-C02-02, FPA2007-63074, and FIS2009-11621-C02-01), by
CPAN-Ingenio (CSPD-2007-00042-Ingenio 2010), by Junta de Andaluc\'{\i}a (FQM160, FQM318, P05-FQM437, and P07-FQM-02962), and by the Universidad Complutense de Madrid (UCM-910059). P. P-F. is supported by a FPU grant of the Spanish Ministerio de Educaci\'on y Ciencia. A. R. is supported by the Spanish program CPAN Consolider-ingenio 2010.

\end{document}